# GScomp-QA: A Subjective Dataset for Quality Assessment of Compressed Gaussian Splatting

Pedro Martin, António Rodrigues, João Ascenso, Maria Paula Queluz
Instituto de Telecomunicações, Instituto Superior Técnico, University of Lisbon, Lisbon, Portugal

*Abstract*—**Gaussian Splatting (GS) has emerged as an efficient representation for high-quality 3D reconstruction and novel view synthesis. However, its large model size poses challenges for storage and transmission. While several GS compression solutions have been proposed, their perceptual impact remains poorly understood due to the lack of dedicated evaluation datasets. To address this gap, this paper introduces GScomp-QA, a subjective quality assessment dataset for evaluating synthesis quality from compressed GS models. The dataset comprises 331 video stimuli from 13 real-world scenes, covering 9 state-of-the-art GS compression solutions. By using videos synthesized from uncompressed models as reference, GScomp-QA tends to isolate compression effect from view synthesis. A subjective study with 20 participants was conducted, providing reliable perceptual scores. Based on these data, GS compression solutions are evaluated through perceptual rate-distortion analysis. In addition, 18 objective quality metrics are evaluated, showing that they do not fully capture GS-specific distortions. GScomp-QA is publicly available and provides a benchmark for evaluating GS compression solutions and supporting the development of quality metrics tailored to GS compression: github.com/pedrogcmartin/GScomp-QA.**

## I. Introduction

The demand for immersive 3D visual experiences has grown significantly in recent years. Applications such as virtual reality, augmented reality, telepresence, and digital heritage require scene representations that provide high-quality rendering from arbitrary viewpoints while maintaining real-time performance. Gaussian Splatting (GS) [1] has emerged as a promising paradigm for novel view synthesis, representing scenes as collections of 3D Gaussians parameterized by spatial attributes and view-dependent appearance. Despite its advantages, the large size of GS models poses significant challenges for storage and transmission, motivating the development of dedicated compression solutions. However, assessing the perceptual impact of GS compression is challenging, as compression artifacts are entangled with distortions inherent to view synthesis. Existing studies often ignore the entanglement between compression and synthesis effects, making it difficult to assess the perceptual impact of compression. Furthermore, the scarcity of publicly available perceptual datasets for compressed GS hinders both the design of dedicated quality metrics and the benchmarking of compression solutions.

To address these limitations, this paper introduces GScomp-QA, a subjective quality assessment dataset for compressed GS view synthesis, aiming to provide a deeper understanding of the perceptual impact of GS compression. The dataset supports three key use cases: *i)* developing perceptually grounded objective quality metrics suitable for GS codec optimization and evaluation; *ii)* benchmarking state-of-the-art GS compression solutions; and *iii)* validating existing and future quality metrics for compressed GS models. The main contributions of this work are as follows:

- **Perceptual dataset for compressed GS view synthesis:** GScomp-QA comprises synthesized videos and subjective scores, obtained from a subjective assessment of several state-of-the-art GS compression solutions across diverse scenes. The use of renderings from uncompressed GS models as references provides a controlled evaluation setting in which perceptual differences associated with compression can be more reliably studied.
- **Benchmarking of GS compression solutions:** A comparative study of GS compression strategies is conducted, providing insights into model-dependent behaviors and perceptual rate-quality trade-offs.
- **Objective quality metrics evaluation:** A comprehensive evaluation of objective quality metrics is conducted against the subjective scores, assessing their reliability and suitability for predicting the perceptual impact of GS compression. The dataset is publicly available at [2] under a CC BY-SA 4.0 license.

## II. Related Work

Quality assessment of GS-generated content has received growing attention in recent years. GS-QA [3] introduced the first subjective dataset evaluating GS view synthesis quality across multiple GS methods and real-world scenes. However, it does not investigate the perceptual impact of GS model compression. 3DGS-IEval-15K [4] is an image dataset specifically designed for compressed GS models. While valuable, it does not fully reflect practical scenarios, where view synthesis is typically experienced as a continuous visual stream through camera motion or interactive exploration. Temporal artifacts such as flickering, geometric instability, and view-dependent inconsistencies are often perceptible only during viewpoint transitions, making them difficult to capture through isolated image evaluation. Furthermore, this dataset is not currently publicly available. 3DGS-VBench [5] evaluates compressed GS models using synthesized videos but is also not publicly accessible. Furthermore, it adopts a categorical rating methodology without an explicit reference sequence, making it unable to isolate the quality impact of compression from view synthesis artifacts. Other GS quality assessment datasets only

This work is funded by FCT through national funds under 2025.01894.BD and, when eligible, co-funded by EU funds under UID/50008/2025 – Instituto de Telecomunicações (DOI: https://doi.org/10.54499/UID/50008/2025).

partially address compression-related distortions. 3DGS-QA [6] and MUGSQA [7] focus on distortions arising from different number of training images, number of training iterations, noise presence, or downsampling, rather than artifacts introduced by dedicated compression solutions. Moreover, 3DGS-QA and MUGSQA are limited to synthetic scenes, which lack the structural complexity and perceptual richness of real-world captures. GSCQA [8] includes compressed GS content, but its scope is limited to the artifacts produced only by their proposed compression pipeline.

The proposed GScomp-QA addresses these limitations: *i)* subjects evaluate synthesized videos rather than isolated images, enabling the assessment of temporal and view-dependent artifacts: *ii)* subjective quality is measured using the uncompressed GS representation as reference, allowing a controlled comparison that reduces the influence of synthesis artifacts; *iii)* it covers two baseline GS representations paired with several compression solutions, ensuring diverse compression effects; and *iv)* the dataset includes real-world scenes, preserving the richness and complexity of natural content. Together, these design choices make GScomp-QA a comprehensive benchmark for GS compression evaluation and objective quality metric development and validation.

## III. GSCOMP-QA DATASET DESIGN

This section describes the framework used to create the proposed dataset, comprising the stimuli and the associated subjective scores.

### A. *Framework Overview*

The GScomp-QA framework takes as input a set of captured images, with the respective camera poses and the point cloud (PC) generated by a structure-from-motion algorithm [9, 10], and produces the test and reference sequences for several GS compression solutions and associated quality scores. As depicted in Fig. 1, the framework has two parallel branches: the reference branch (top) synthesizes video from an uncompressed (baseline) GS model, while the test branch (bottom) synthesizes video from a compressed GS model. The uncompressed baseline serves as reference, instead of the original captured content, allowing the compression effect to be assessed with a smaller influence of view synthesis artifacts. The GScomp-QA dataset comprises the synthesized videos and their associated subjective scores (red box in Fig. 1), generated through these stages:

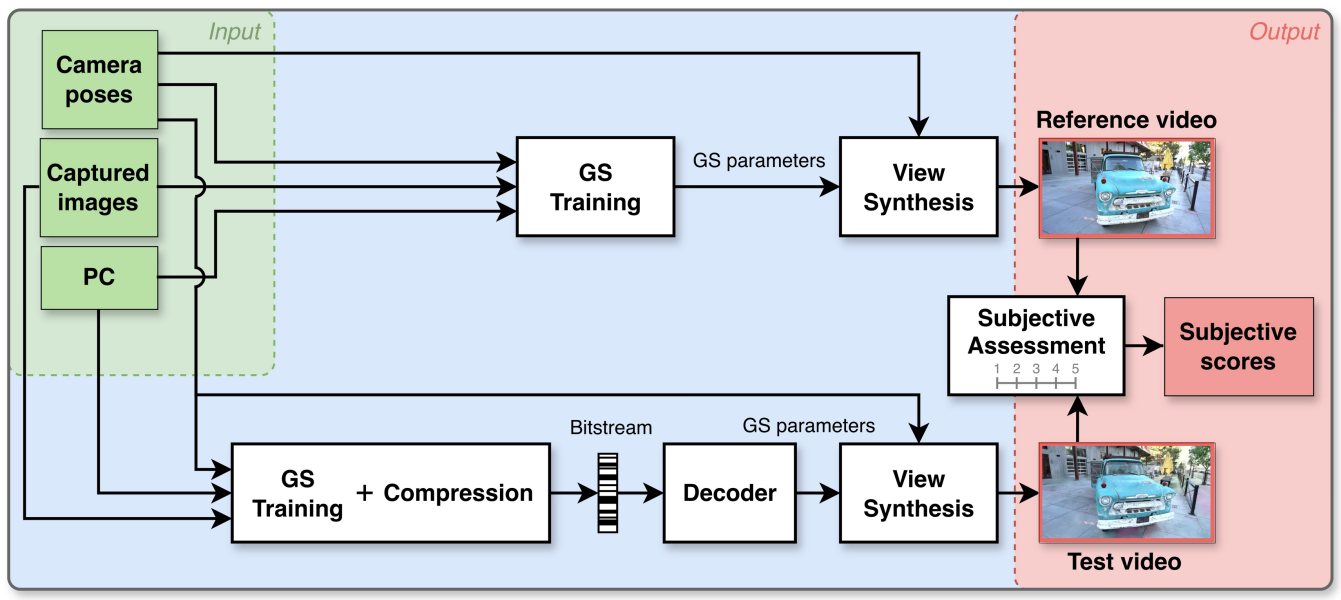

Fig. 1: GScomp-QA framework.

- **GS Training:** A baseline GS model is trained using the input images, PC, and camera poses (green box in Fig. 1). During training the model parameters are optimized end-to-end by minimizing a reconstruction loss between synthesized and ground-truth training views. Two GS baselines are considered: 3DGS [1], which represents Gaussian parameters explicitly, and Scaffold-GS [11], which encodes them implicitly through voxel-grid features and MLPs, thus offering two fundamentally different scene representations. Existing GS compression solutions are typically designed for one of these two representations. In 3DGS, each Gaussian is independently defined by explicit parameters including position, scale, rotation, opacity, and view-dependent color. Scaffold-GS instead anchors the representation on a sparse voxel grid, where local Gaussian attributes are predicted using MLPs.
- **GS Training plus Compression:** Compression may be coupled with the training process (model creation) or applied as a post-training stage. In both cases, the GS model parameters are encoded into a compact bitstream. Section III.B details the selected compression approaches.
- **Decoder:** The bitstream is decoded to reconstruct the GS parameters, which are then used for view synthesis. The decoder mirrors the inverse operations of the encoding process, recovering the GS model.
- **View Synthesis:** Novel views are synthesized from both the compressed and uncompressed GS models, producing the test and reference videos, respectively. For each input scene, all videos are synthesized using an identical camera trajectory to ensure a valid comparison. The trajectory is derived from the training camera poses, following the approach of [12]: a focal point is estimated as the point that minimizes the sum of squared distances to all training camera optical axes; an elliptical orbit is then fitted around this focal point, with semi-axes set to the 90th percentile of the training camera position deviations along each spatial dimension. A 10s video with 24 fps is then synthesized along this trajectory.
- **Subjective Assessment:** The resulting video pairs are evaluated with a semi-controlled subjective assessment experiment, described in Section IV.B. The reference video of the pair is synthesized from an uncompressed baseline GS model (either 3DGS or Scaffold-GS), while each test video of the pair is synthesized from a compressed model produced by a codec designed for that baseline.

### B. *Selected GS Compression Solutions*

This section presents the GS compression solutions selected for the creation of the dataset, broadly categorized into two classes based on their interaction with the training process.

*1) Joint Training and Compression:* These methods integrate compression constraints directly into the GS training process, jointly optimizing for both reconstruction quality and model compactness. This class is also denominated as A-3DGS (Alternative 3DGS) according to MPEG taxonomy [13]. For 3DGS-based representations, Reduced3DGS [14], Compact3D [15], and LightGaussian [16] progressively prune redundant Gaussians during training and apply vector quantization to the parameters of the retained Gaussians. SOG [17] modifies the GS training process itself to organize Gaussian parameters into a locally smooth 2D grid, making the resulting model more amenable to image-based compression. For Scaffold-GS-based representations, HAC++ [18] exploits the relationship between unorganized anchors and a structured hash grid, using their mutual information for context modeling, combined with intra-anchor redundancy reduction. ContextGS [19] introduces an autoregressive context model that organizes anchors into hierarchical levels, where already coded anchors at coarser levels are used to predict finer-level anchors, improving entropy coding efficiency. Joint training and compression solutions generally achieve high rate-distortion (RD) performances, as the model learning is explicitly optimized for compression. However, since compression is tightly coupled with the training process, these methods require access to the original training data, which is rarely available in practical deployment scenarios. Regarding quality evaluation, this training-compression entanglement means that the evaluated perceptual differences cannot be attributed only to compression. Nevertheless, comparison with the corresponding baseline provides a controlled setting for evaluating the perceptual changes associated with the compressed representation.

*2) Post-Training Compression:* These methods compress pre-trained GS models directly without requiring any GS retraining. According to the MPEG taxonomy, this approach is referred as I-3DGS (Inria 3DGS) [13]. For 3DGS-based representations, GSICO-3DGS [20] maps Gaussian parameters onto structured 2D images through a spatially coherent arrangement algorithm, enabling their compression with a conventional image codec; FCGS [21] employs a learned feed-forward compression model with context modeling, enabling compression of any pre-trained model in a single pass without per-scene optimization. GSICO-Scaffold-GS [20] extends the image-based coding approach to the voxel-based representation of Scaffold-GS, and is, to the best of the authors' knowledge, the only post-training compression solution specifically designed for this representation.

### *C. Selected Visual Scenes*

The GScomp-QA dataset is built upon visual scenes drawn from publicly available datasets widely used in GS research, selected to ensure diversity in scene structure, scale, and capture conditions: Tanks and Temples (T&T) [22], with *train* and *truck* scenes; Deep Blending (DB) [23], with *drjohnson* and *playroom*; Light Field (LF) [24], with *africa*, *basket*, and *ship*; and MipNeRF360 (M360) [25], with *bicycle*, *bonsai*, *counter*, *flowers*, *garden*, and *kitchen*; resulting in a total of 13 scenes. These are static scenes that span a wide variety of characteristics: indoor and outdoor scenes, 360° and freestyle camera acquisition, small-scale environments to large-scale settings. Table I specifies the number of training images and spatial resolution of each scene.

## IV. Subjective Test Design and Procedure

This section describes the subjective test design and procedure used to generate the GScomp-QA dataset.

TABLE I. SELECTED VISUAL SCENES SPECIFICATIONS

| Visual scene | #Train. images | Resolution [px.] | Visual scene | #Train. images | Resolution [px.] |
|---|---|---|---|---|---|
| *africa* | 56 | 1269×709 | *garden* | 161 | 1297×840 |
| *basket* | 74 | 1293×721 | *kitchen* | 244 | 1558×1039 |
| *bicycle* | 169 | 1237×822 | *playroom* | 196 | 1264×832 |
| *bonsai* | 255 | 1559×1039 | *ship* | 95 | 1266×711 |
| *counter* | 210 | 1558×1038 | *train* | 263 | 980×545 |
| *drjohnson* | 230 | 1332×876 | *truck* | 219 | 979×546 |
| *flowers* | 151 | 1256×828 | – | – | – |

### *A. Test Conditions*

Each compression solution has one or more parameters to control the trade-off between model size (rate) and synthesis quality. For ContextGS, FCGS, and HAC++, this trade-off is explicitly controlled by a Lagrangian multiplier, $\lambda$, within the loss function. Compact3D and Reduced3DGS adjust the rate through VQ codebook sizes, with Reduced3DGS also using a Gaussian pruning level. SOG and GSICO-3DGS control compression through the JPEG XL quality level, while GSICO-Scaffold-GS uses the quantization step size of the anchor and offset parameters. LightGaussian controls compression through Gaussian pruning percentages.

For each method, three operating points were defined: high quality (HQ), medium quality (MQ), and low quality (LQ), corresponding to progressively higher perceptual degradation (and lower rates). These operating points were determined through an expert pre-screening session, in which synthesized videos produced under several parameter settings were visually inspected for each method and scene, and the settings yielding the desired quality for each operating point were selected. The selected parameter settings for all scenes and compression solutions are provided on the dataset webpage. All GS models were trained and compressed on two GeForce RTX 4090 GPUs.

Across all compression solutions, scenes, and quality levels, a total of 331 stimuli were generated for subjective evaluation. This includes 26 reference sequences obtained by the two uncompressed baseline models for 13 scenes, and 305 test sequences corresponding to the compressed versions. While most methods were evaluated at the three quality levels across all 13 scenes, there are a few exceptions due to inherent limitations of some methods. FCGS was evaluated only at HQ, as its context model is exclusively trained for HQ compression. SOG was evaluated at HQ and MQ only, as even at its highest compression setting perceptual degradation was not very perceptible and scene geometry remained well-preserved. The same limitation was observed for GSICO-3DGS in 4 scenes (*counter*, *kitchen*, *flowers*, and *garden*). Finally, GSICO-Scaffold-GS does not provide an MQ level for the LF scenes, as quality degradation is abrupt when rate changes.

### *B. Subjective Methodology*

The subjective assessment study follows the DSCQS methodology as defined in ITU-R BT.500 [26]. DSCQS is particularly suitable for GS compression evaluation because participants rate both videos independently on a continuous scale, and without knowing which one is the reference. Moreover, this naturally accommodates cases where, due to compression, the test video is perceptually preferred over the uncompressed reference [27]. In each trial, participants are simultaneously shown the test and reference videos side-by-side, and rate each independently using a continuous slider with five quality levels, Bad, Poor, Fair, Good, and Excellent, corresponding to a scale from 1 to 5. The spatial position of the reference video and the presentation order of trials are randomized to minimize positional and contextual bias. Each video loops three times and participants can submit their scores only after the first playback.

To mitigate participant fatigue, the study is organized into three balanced test sessions, each covering a representative subset of compression solutions, scenes, and quality levels. For each scene, pairs of reference videos obtained from the two uncompressed baseline models were included, to enable score alignment and fair comparison between compression solutions using different baselines (cf. Section V.B), allowing all DMOS values to be expressed on a common quality scale. The test is conducted on an ASUS ProArt PA32UC-K 4K HDR monitor at a spatial resolution of 1920×1080 pixels. To enable side-by-side presentation within the available screen space, each video is downsampled to a width of 940 pixels using bicubic interpolation, preserving the original aspect ratio.

Twenty non-expert observers participated in the study, 16 males and 4 females, aged between 18 and 35. Before the test, participants were informed of the test objective and completed a training session to familiarize themselves with the evaluation procedure and the range of expected visual distortions.

### C. *Data Processing*

After data collection, outlier participants were identified using the kurtosis-based and correlation-based screening procedures recommended by ITU-R BT.500 [26], applied to the participants' raw opinion scores, resulting in the exclusion of two participants by the second one; thus, $N = 18$ valid participants were retained. The opinion scores assigned to each pair of videos by each participant $p \in \{1, \ldots, N\}$, for compression solution $i$ with quality level $l$, against baseline model $j$ and scene $k$, are denoted $\mathrm{OS}_{\mathrm{test}}^{(p,i,k,l)}$ and $\mathrm{OS}_{\mathrm{ref}}^{(p,j,k)}$, respectively. The differential opinion score (DOS) of a participant p for a given pair of videos is defined as:

$$\mathrm{DOS}^{(p,i,j,k,l)} = \mathrm{OS}_{\mathrm{test}}^{(p,i,k,l)} - \mathrm{OS}_{\mathrm{ref}}^{(p,j,k)} \tag{1}$$

and the differential mean opinion score (DMOS) was obtained by averaging DOS across all valid participants:

$$\mathrm{DMOS}^{(i,j,k,l)} = 5 + \frac{1}{N}\sum_{p=1}^{N} \mathrm{DOS}^{(p,i,j,k,l)} \tag{2}$$

Under this formulation, higher DMOS values indicate higher perceived quality, preserving the intuitive interpretation of the MOS scale. Notably, DMOS values exceeding the nominal MOS range (i.e., 5) may occur when the test sequence was judged perceptually superior in quality to the baseline reference.

When comparing compression solutions designed for different baseline models, the DMOS values computed using (2) are not directly comparable, as they are expressed relative to different references, namely videos synthesized from different uncompressed GS models. To enable a fair unified comparison, a reference compensation is applied to adapt the DMOS of Scaffold-GS-based solutions relative to 3DGS, which was chosen as the common reference given it is the seminal model. Reference compensation is performed using the direct comparison between the videos that resulted from the two uncompressed baseline models. For each participant $p$, Scaffold-GS-based compression solution $i$, scene $k$, and quality level $l$, the compensated DOS with respect to 3DGS as reference, is derived as:

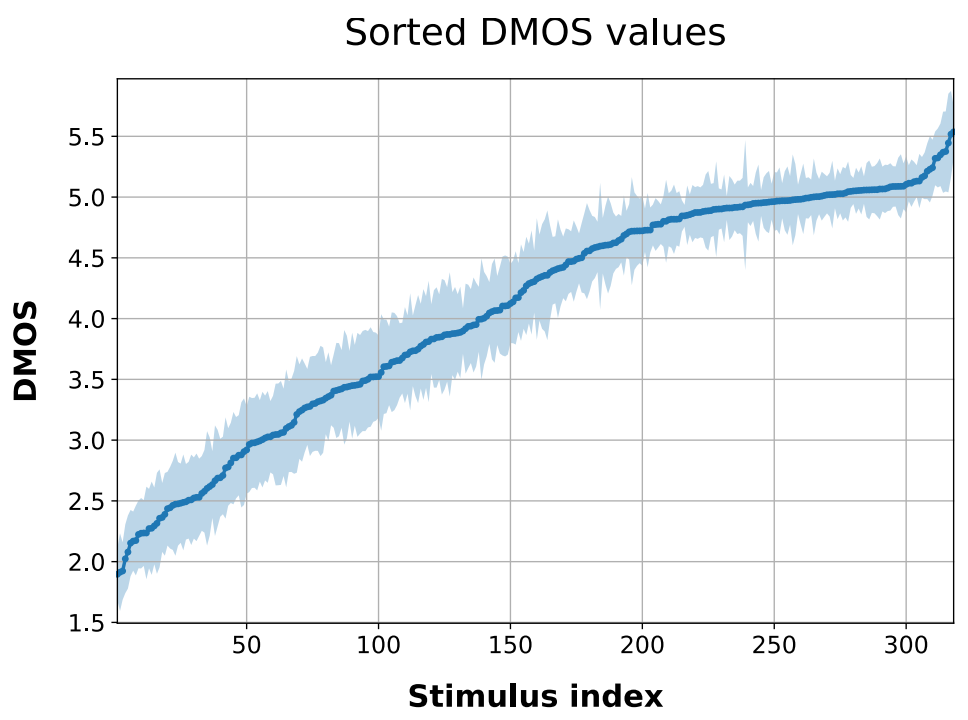

Fig. 2. Sorted DMOS values with 95% confidence intervals.

$$\mathrm{DOS}_{\mathrm{comp}}^{(p,i,3\mathrm{DGS},k,l)} = \mathrm{OS}_{\mathrm{test}}^{(p,i,k,l)} - \mathrm{OS}_{\mathrm{ref}}^{(p,3\mathrm{DGS},k)}$$
$$= \left(\mathrm{OS}_{\mathrm{test}}^{(p,i,k,l)} - \mathrm{OS}_{\mathrm{ref}}^{(p,\mathrm{SGS},k)}\right) - \left(\mathrm{OS}_{\mathrm{ref}}^{(p,3\mathrm{DGS},k)} - \mathrm{OS}_{\mathrm{ref}}^{(p,\mathrm{SGS},k)}\right) \tag{3}$$
$$= \mathrm{DOS}^{(p,i,\mathrm{SGS},k,l)} - \mathrm{DOS}^{(p,3\mathrm{DGS},\mathrm{SGS},k)}$$

where $\mathrm{OS}_{\mathrm{ref}}^{(p,3\mathrm{DGS},k)}$ and $\mathrm{OS}_{\mathrm{ref}}^{(p,\mathrm{SGS},k)}$ are the participant's opinion score regarding the 3DGS and Scaffold-GS (or SGS) baseline models, respectively, $\mathrm{DOS}^{(p,i,\mathrm{SGS},k,l)}$ is the DOS between the Scaffold-GS-based compression solution and Scaffold-GS, and $\mathrm{DOS}^{(p,3\mathrm{DGS},\mathrm{SGS},k)}$ is the DOS between the two baseline models. The other equation elements are as defined in (1). The compensated DMOS is then obtained by averaging the compensated DOS over all participants and adding an offset of 5. This procedure enables a unified RD comparison of all compression solutions regardless of their GS baseline model.

## V. Experimental Setup and Analysis

This section analyzes the subjective scores collected in the experiment and benchmarks the considered GS compression solutions. Section V.A presents and analyzes the DMOS scores, including a non-inferiority statistical test. Section V.B evaluates compression performance through perceptual RD analysis.

### A. *Quality Assessment*

Fig. 2 presents the DMOS values of all tested stimuli, sorted in ascending order, together with their 95% confidence intervals (CI). The scores span a wide perceptual range, from low to high quality, confirming that the selected compression configurations provide a representative and diverse set of perceptual conditions. A Cronbach's alpha of 0.983 was obtained across all participant scores, indicating excellent inter-participant agreement and validating the reliability of the collected subjective data.

A significant number of stimuli exhibit DMOS values greater than 5, indicating that compressed versions are perceived as having higher quality than the uncompressed reference. This effect is particularly pronounced for joint training and compression solutions, where compression integrated into training acts as regularization, reducing noise and redundancy in the Gaussian representation. Methods such as ContextGS and HAC++ exploit spatial correlations through context modeling and structured representations, promoting consistency across neighboring Gaussians. This reduces geometric instability and enables more coherent rendering, especially in challenging regions. Similar effects are observed in post-training methods where both pruning and learned compression can act as implicit regularizers of the GS representation.

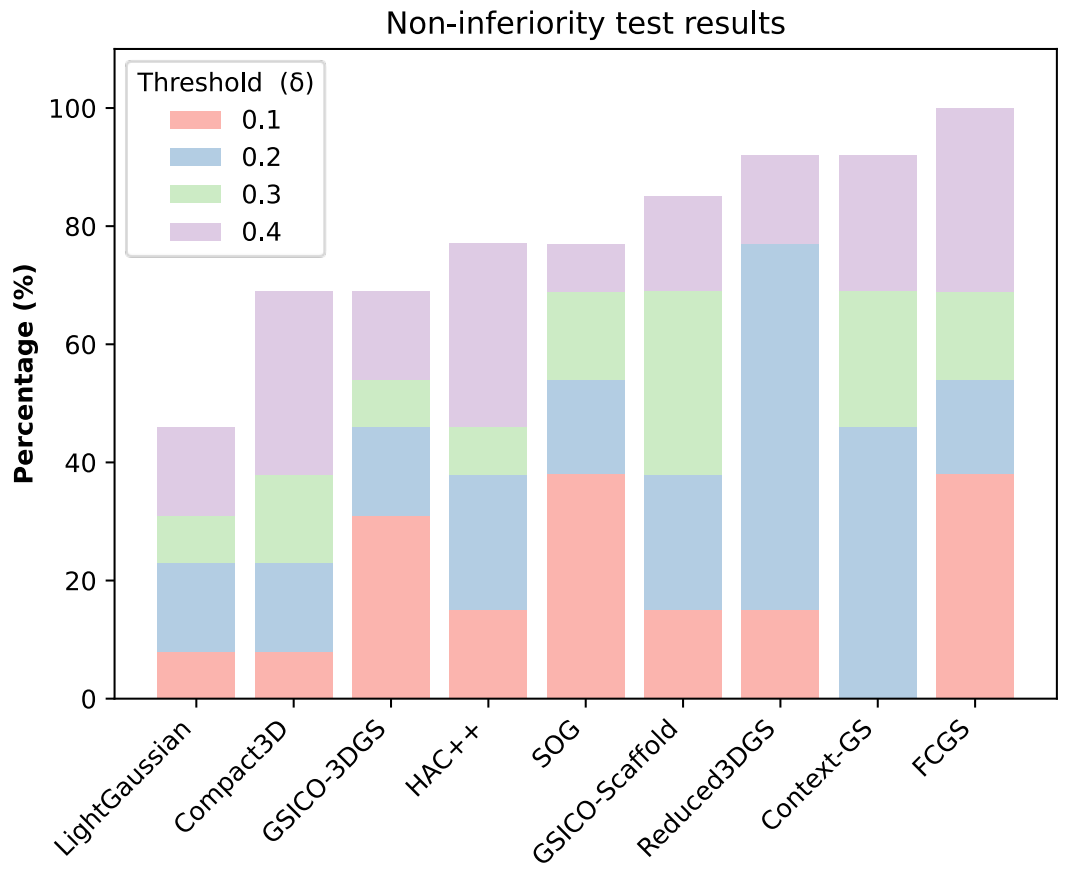

Fig. 3. Non-inferiority statistical test results per GS compression solution

To formally assess cases where compressed sequences are perceptually equivalent or superior to the uncompressed reference, a non-inferiority test is applied to HQ stimuli. The null hypothesis states that the quality of the compressed sequence is inferior to the reference by more than a predefined threshold $\delta$. Rejection of the null hypothesis occurs when the lower bound of the 95% confidence interval of the difference between the mean opinion scores of the compressed and reference sequences exceeds -$\delta$, indicating that the compressed sequence is not perceptually worse than the reference beyond the accepted tolerance. In terms of DMOS, this is equivalent to requiring the lower bound of the 95% CI to exceed $5 - \delta$. The test was applied for $\delta \in \{0.1, 0.2, 0.3, 0.4\}$. Fig. 3 shows, for each compression solution, the percentage of HQ stimuli for which the null hypothesis is rejected at each $\delta$. Full statistical results are available on the GScomp-QA dataset webpage. Fig. 3 shows that most compression solutions are able to achieve perceptual quality levels equivalent to or better than the reference. In particular, joint training and compression solutions, such as Reduced3DGS and ContextGS, rank among the best-performing methods, with 92% of HQ compressed sequences generated by those methods not perceived as inferior to the reference at a threshold of 0.4.

### *B. GS Compression*

RD performance is evaluated by defining the rate as the bitstream size in megabytes (MB) and the distortion as the average DMOS of each scene and quality level. Fig. 4 depicts RD curves (with the respective DMOS CIs) for four representative scenes, one from each dataset. RD plots for the remaining scenes are provided in the dataset webpage [2]. For Scaffold-GS-based compression solutions, the compensated

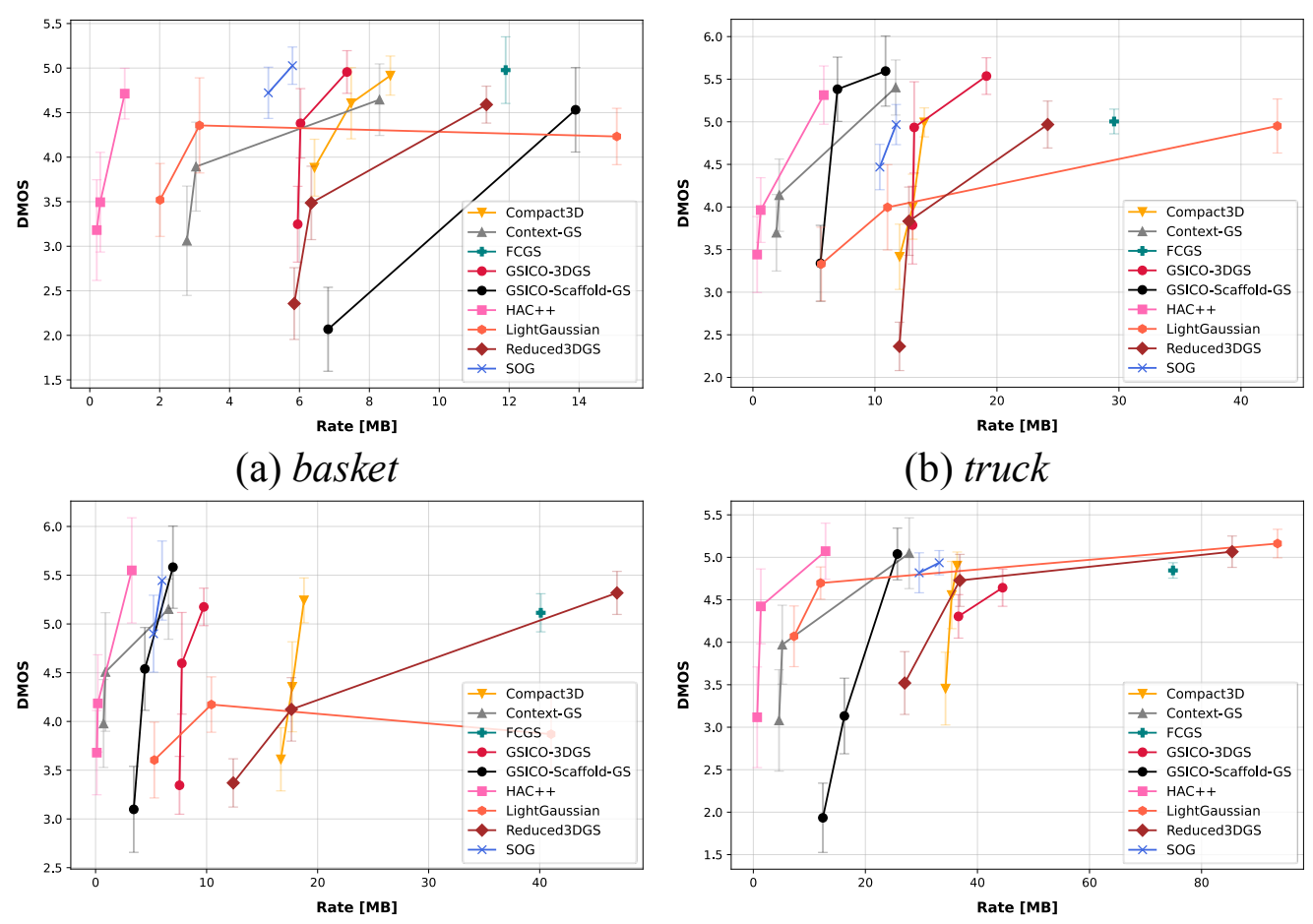

(a) *basket* (b) *truck*

(c) *drjohnson* (d) *garden*

Fig. 4. RD curves for benchmarking GS compression.

DMOS values are used. These results reveal clear trends across compression strategies. Joint training and compression solutions, particularly HAC++ and ContextGS, are the best at low rates. This stems from jointly optimizing representation and compression, allowing them to exploit strong spatial correlations and structural priors more effectively. LightGaussian also demonstrates the benefits of combining Gaussian pruning with model fine-tuning to jointly reduce model complexity and representation noise. Post-training methods exhibit lower compression efficiency, as they operate on fixed representations. Among post-training approaches, the image-based approach adopted by GSICO-Scaffold-GS achieves the best RD trade-offs, highlighting the effectiveness of leveraging mature 2D coding techniques for GS compression.

## VI. Objective Quality Metrics Benchmarking

This section evaluates how well existing objective quality metrics predict the perceptual impact of GS compression, using the GScomp-QA dataset. A broad set of full-reference metrics is considered (following the selection proposed in [3]), including classical solutions, namely FSIM, FVVDP, GMSD, IW-SSIM, MAD, MSE-RGB, MS-SSIM, NLPD, PSNR-HVS, PSNR-Y, PSNR-YUV, SSIM, VIF, and VSI, as well as learning-based approaches, namely DISTS, LPIPS, ST-LPIPS, and VMAF. The performance of each metric is assessed by measuring its correlation with subjective scores, using three standard criteria: the Pearson Linear Correlation Coefficient (PLCC), computed after cubic regression; Spearman Rank Order Correlation Coefficient (SROCC); Root Mean Squared Error (RMSE); and Perceptually Weighted Ranking Correlation (PWRC) [28]. Experimental results are summarized in Table II. RMSE and PWRC results are presented in the dataset webpage [3], alongside a statistical significance analysis of the metrics' performances.

Overall, NLPD achieves the highest correlation across all criteria (with PLCC of 0.88 and SROCC of 0.87), followed by ST-LPIPS, FSIM, GMSD, and VMAF, reflecting the benefits of incorporating structural sensitivity and visual system modeling or adopting a learning-based approach. However, even the best-performing metrics fail to achieve near-perfect correlation. This limitation stems from the specific nature of GS compression artifacts, which differ from those in conventional image and video coding. In particular, GS compression introduces geometry-related distortions, view-dependent inconsistencies, and temporal instabilities that are not explicitly modeled by existing metrics. As a result, current metrics may not reliably guide the optimization of GS compression solutions, highlighting the need for specific metrics, positioning GScomp-QA as a benchmark for their development and validation.

TABLE II. CORRELATION RESULTS OF OBJECTIVE QUALITY METRICS

| Metric | PLCC | SROCC | Metric | PLCC | SROCC |
|---|---|---|---|---|---|
| MSE-RGB | 0.593 | 0.621 | VSI | 0.726 | 0.713 |
| PSNR-Y | 0.610 | 0.611 | MAD | 0.735 | 0.726 |
| PSNR-YUV | 0.558 | 0.567 | LPIPS | 0.738 | 0.732 |
| PSNR-HVS | 0.729 | 0.716 | ST-LPIPS | 0.822 | 0.830 |
| SSIM | 0.646 | 0.636 | DISTS | 0.673 | 0.668 |
| MS-SSIM | 0.643 | 0.641 | GMSD | 0.825 | 0.823 |
| IW-SSIM | 0.796 | 0.778 | NLPD | 0.881 | 0.866 |
| VIFp | 0.791 | 0.778 | VMAF | 0.825 | 0.800 |
| FSIM | 0.830 | 0.816 | FVVDP | 0.678 | 0.668 |

## VII. CONCLUSION

This paper proposes GScomp-QA, a subjective dataset for the quality assessment of compressed GS view synthesis, assessing the perceptual impact of compression relative to a consistent GS rendering baseline. Based on this dataset, a benchmark of recent GS compression solutions was conducted, showing that joint training and compression approaches achieve the best perceptual rate–distortion performance. A set of commonly used objective quality metrics was further evaluated, showing that while perceptually motivated metrics such as NLPD achieve higher correlation with subjective scores, existing metrics remain insufficient to fully capture GS-compression specific distortions. These results highlight the need for dedicated metrics capable of guiding the optimization and evaluation of GS compression solutions. GScomp-QA provides a resource to support this research direction.

## REFERENCES

[1] B. Kerbl, G. Kopanas, T. Leimkuehler, and G. Drettakis, "3D Gaussian Splatting for Real-Time Radiance Field Rendering," *ACM Trans. Graph.*, vol. 42, no. 4, Jul. 2023.

[2] P. Martin, A. Rodrigues, J. Ascenso, and M. P. Queluz, "GScomp-QA: A Subjective Dataset for Quality Assessment of Compressed Gaussian Splatting," GitHub repository, 2026. [Online]. Available: https://github.com/pedrogcmartin/GScomp-QA.

[3] P. Martin, A. Rodrigues, J. Ascenso, and M. P. Queluz, "GS-QA: Comprehensive Quality Assessment Benchmark for Gaussian Splatting View Synthesis," in *Proc. 17th IEEE Int. Conf. Quality Multimedia Experience (QoMEX)*, 2025, pp. 1–7.

[4] Y. Xing, J. Wang, P. Niu, W. Huang, G. Zhai, and Y. Xu, "3DGS-IEval-15K: A Large-scale Image Quality Evaluation Database for 3D Gaussian-Splatting," in *Proc. 33rd ACM Int. Conf. Multimedia (MM)*, 2025, pp. 12682–12689.

[5] Y. Xing, W. Gordon, Q. Yang, K. Yang, J. Wang, and Y. Xu, "3DGS-VBench: A Comprehensive Video Quality Evaluation Benchmark for 3DGS Compression," in *Proc. IEEE Int. Conf. Visual Communications and Image Processing (VCIP)*, 2025, pp. 1–5.

[6] Z. Wan, Y. Diao, J. Xu, H. Wang, Z. Li, X. Fan, W. Zuo, and D. Zhao, "Perceptual Quality Assessment of 3D Gaussian Splatting: A Subjective Dataset and Prediction Metric," in *Proc. AAAI Conf. Artif. Intell. (AAAI)*, vol. 40, no. 12, 2026, pp. 9657–9665.

[7] T. Chen, J. Jin, S. Cai, Z. Li, and W. Lin, "MUGSQA: Novel Multi-Uncertainty-Based Gaussian Splatting Quality Assessment Method, Dataset, and Benchmarks," *arXiv:2511.06830 [cs.CV]*, 2026.

[8] Q. Yang, K. Yang, Y. Xing, Y. Xu, and Z. Li, "A Benchmark for Gaussian Splatting Compression and Quality Assessment Study," in *Proc. 6th ACM Int. Conf. Multimedia Asia (MMAsia)*, 2024, p. 12.

[9] J. L. Schönberger and J.-M. Frahm, "Structure-from-Motion Revisited," in *Proc. IEEE/CVF Conf. Comput. Vis. Pattern Recognit. (CVPR)*, 2016.

[10] J. L. Schönberger, E. Zheng, M. Pollefeys, and J.-M. Frahm, "Pixelwise View Selection for Unstructured Multi-View Stereo," in *Proc. Eur. Conf. Comput. Vis. (ECCV)*, 2016.

[11] T. Lu, M. Yu, L. Xu, Y. Xiangli, L. Wang, D. Lin, and B. Dai, "Scaffold-GS: Structured 3D Gaussians for View-Adaptive Rendering," in *Proc. IEEE/CVF Conf. Comput. Vis. Pattern Recognit. (CVPR)*, 2024, pp. 20654–20664.

[12] S. Girish, K. Gupta, and A. Shrivastava, "Eagles: Efficient Accelerated 3D Gaussians with Lightweight Encodings," in *Proc. Eur. Conf. Comput. Vis. (ECCV)*, 2024, pp. 54–71.

[13] Y. Liao, "Towards Efficient Volumetric Video: Compression and Standardization of Gaussian Splats," *JVIP Webinar Series*, 2025.

[14] P. Papantonakis, G. Kopanas, B. Kerbl, A. Lanvin, and G. Drettakis, "Reducing the Memory Footprint of 3D Gaussian Splatting," *ACM Comput. Graph. Interact. Tech.*, vol. 7, no. 1, 2024.

[15] K. L. Navaneet, K. P. Meibodi, S. A. Koohpayegani, and H. Pirsiavash, "Compact3D: Smaller and Faster Gaussian Splatting with Vector Quantization," *arXiv:2311.18159 [cs.CV]*, 2023.

[16] Z. Fan, K. Wang, K. Wen, Z. Zhu, D. Xu, and Z. Wang, "LightGaussian: Unbounded 3D Gaussian Compression with 15× Reduction and 200+ FPS," in *Proc. Adv. Neural Inf. Process. Syst. (NeurIPS)*, Vancouver, BC, Canada, 2024.

[17] W. Morgenstern, F. Barthel, A. Hilsmann, and P. Eisert, "Compact 3D Scene Representation via Self-Organizing Gaussian Grids," in *Proc. Eur. Conf. Comput. Vis. (ECCV)*, Cham, Switzerland: Springer Nature Switzerland, 2025, pp. 18–34.

[18] Y. Chen, Q. Wu, W. Lin, M. Harandi, and J. Cai, "HAC++: Towards 100X Compression of 3D Gaussian Splatting," *arXiv:2501.12255 [cs.CV]*, 2025.

[19] Y. Wang, Z. Li, L. Guo, W. Yang, A. C. Kot, and B. Wen, "ContextGS: Compact 3D Gaussian Splatting with Anchor Level Context Model," *arXiv:2405.20721 [cs.CV]*, 2024.

[20] P. Martin, A. Rodrigues, J. Ascenso, and M. P. Queluz, "Structured Image-based Coding for Efficient Gaussian Splatting Compression," *arXiv:2601.14510 [cs.MM]*, 2026.

[21] Y. Chen, Q. Wu, M. Li, W. Lin, M. Harandi, and J. Cai, "Fast Feedforward 3D Gaussian Splatting Compression," *arXiv:2410.08017 [cs.CV]*, 2025.

[22] A. Knapitsch, J. Park, Q.-Y. Zhou, and V. Koltun, "Tanks and Temples: Benchmarking Large-Scale Scene Reconstruction," *ACM Trans. Graph.*, vol. 36, no. 4, pp. 1–13, 2017.

[23] P. Hedman, J. Philip, T. Price, J.-M. Frahm, G. Drettakis, and G. Brostow, "Deep Blending for Free-Viewpoint Image-Based Rendering," *ACM Trans. Graph.*, vol. 37, no. 6, pp. 1–15, 2018.

[24] K. Yucer, A. Sorkine-Hornung, O. Wang, and O. Sorkine-Hornung, "Efficient 3D Object Segmentation from Densely Sampled Light Fields with Applications to 3D Reconstruction," *ACM Trans. Graph.*, vol. 35, no. 3, 2016.

[25] J. T. Barron, B. Mildenhall, D. Verbin, P. P. Srinivasan, and P. Hedman, "Mip-NeRF 360: Unbounded Anti-Aliased Neural Radiance Fields," in *Proc. IEEE/CVF Conf. Comput. Vis. Pattern Recognit. (CVPR)*, 2022, pp. 5470–5479.

[26] ITU-R Recommendation BT.500-15, "Methodologies for the Subjective Assessment of the Quality of Television Images," *Int. Telecommun. Union*, Geneva, Switzerland, 2023.

[27] M. T. Bagdasarian, P. Knoll, Y. Li, F. Barthel, A. Hilsmann, P. Eisert, and W. Morgenstern, "3DGS.zip: A Survey on 3D Gaussian Splatting Compression Methods," *Comput. Graph. Forum*, vol. 44, no. 2, Art. no. e70078, 2025.

[28] Q. Wu, H. Li, F. Meng, and K. N. Ngan, "A Perceptually Weighted Rank Correlation Indicator for Objective Image Quality Assessment," *IEEE Trans. Image Process.*, vol. 27, no. 5, pp. 2499–2513, 2018.